\documentclass[preprint,double-spaced,showpacs,amsmath,amssymb]{revtex4}\textwidth=14.5cm \textheight=24cm
\usepackage{amssymb,graphicx}
\usepackage{dcolumn}% Align table columns on decimal point

\def\P3{{\cal P}_t}
\def\J3{{\cal J}}
\def\T3{{\cal T}}

\def\beq{\begin{equation}}
\def\eeq{\end{equation}}
\def\bar{\begin{array}[b]}
\def\barc{\begin{array}}
\def\bart{\begin{array}[t]}
\def\ear{\end{array}}

 1  1
 1 
scaled\magstephalf 
 
scaled\magstephalf

\begin{document}
\thispagestyle{empty}
\vspace*{0.5 cm}
%\vspace*{-1.2in}
%\vspace*{0.7 in}
\begin{center}
{\bf Probing protoneutron star density profile from neutrino signals.}
\\
\vspace*{1cm} {M. Baldo and V. Palmisano}\\
\vspace*{.3cm}
{\it INFN, Sezione di Catania}\\
{\it Via S. Sofia 64, I-95123, Catania, Italy} \\
\vspace*{.6cm}
%\vspace*{.6cm} {\bf }\\ \vspace*{.3cm}
%{\it INFN, Sezione di Catania}\\
%{\it Dipartimento di Fisica, Universit\`a di Catania}\\
%{\it Corso Italia 57, I-95129, Catania, Italy} \\
\vspace*{1 cm}
\end{center}

\begin{center}{\bf ABSTRACT} \\\end{center}

Supernovae of Type II is a phenomenon that occurs at the end of evolution of massive stars when the iron core of
the star exceeds a mass limit. After collapse of the core under gravity the shock wave alone does not succeed in
expelling the mass of the star and in this sense the role of neutrinos is the most important mechanism to do so.
During the emission of neutrinos flavor conversion is possible, related the phenomenon of oscillations, which
however depends directly on the particular density profile of the medium. In this paper we present results of
numerical simulations of neutrino flavor conversion in protoneutron stars and after collapse. The probabilities
of survival for a given flavor in a complete three-flavors framework is estimated through an algorithm which
conserves unitarity to a high degree of accuracy. The sensitivity of the results to the different adopted models
for the protoneutron star structure is examined in detail demonstrating how the neutrino signal could be used to
check the validity of models.

\vskip 0.3 cm
%\noindent

%\vfill\eject

\section{Introduction}

The solution of the solar neutrino "puzzle" \cite{egu,hax} has
pointed out the relevance of the MSW mechanism \cite{MSW} in the
neutrino propagation through the dense medium of astrophysical
objects. The mechanism is a consequence of the neutrino masses and
the corresponding flavor mixing. If neutrino are emitted from the
core of a star, the initial flavor distribution can be distorted by
the flavor conversion that can occur during the propagation
outwards. On one hand this can be a complication in interpreting the
possible neutrino signals, on the other hand the presence of
conversion can be of great value in constraining the models of the
astrophysical objects. The solar neutrino problem is a typical case
where flavor conversion offers a solution of the puzzle and at the
same time put serious constrains on the modeling of the internal
structure of the star, which would not be easily accessible
otherwise. All that has stimulated a large number of studies of
neutrino propagation in more compact stars, like supernovae,
protoneutron stars and neutron stars \cite{bal}. These astrophysical
objects have a much larger central density and therefore neutrino
flavor conversion can occur only in the outer part, where the
density can decrease to values compatible with the MSW
mechanism.\par Of particular interest is the case of protoneutron
stars, where the neutrinos of all three flavors can be considered in
thermal equilibrium and thermally generated at the center of the
stellar object. In principle the neutrino energy spectrum of each
flavor is determined by the temperature at which they decouple from
matter, i.e. the extension of the corresponding neutrino sphere.
However, due to possible conversion through the MSW mechanism, the
energy spectrum and each flavor fraction is modified after the
propagation through the stellar medium. This modification depends
crucially on the density profile, and therefore the neutrino signal
can be used in principle to constraint the models of the
protoneutron star structure and evolution. The effect of the shock
wave on the neutrino signal has been analyzed in several papers
\cite{fuller,fogli,gian}, both as a function of time and in relation
to neutrino mixing  parameters. The dependence on the structure of
the progenitor star has been analyzed in ref. \cite{prog}, while the
expected relic neutrinos characteristics have been studied in ref.
\cite{volpe}.

\par

However, to the authors knowledge, no systematic study of the relation between
possible neutrino signals and protoneutron star structure have been carried out
in the literature. In this work we present a detailed analysis of the neutrino
propagation in the protoneutron star medium and we try to relate directly some
characteristics of the neutrino signals to the star density profile, which, in
principle, could be used to distinguish among different models or model
parameters.\par In this study we will neglect neutrino--neutrino interaction,
which has been recently shown \cite{fuller} to be able to produce giant
coherent flavor conversion. This coherent process can occur under particular
conditions, and, to simplify the analysis, we do not consider this possibility
in the paper.

\section{Generalities and numerical method.}

It is customary to schematize the propagation of neutrinos outside the neutrino sphere as coherent and subject
only to a single particle potential, i.e. only forward scattering is present \cite{lunar}. The medium has then
only the effect of shifting the neutrino masses locally at each point of the star. As it is well known, due to
the dominance of electrons in the lepton content of the protoneutron star, this shift is flavor dependent and
the neutrino mass spectrum is strongly altered. In fact, due to the electron dominance, electron neutrinos
interact also through charged weak current, while the $\tau$ and $\mu$ neutrinos interact essentially through
neutral weak current only. In the deep interior of the star the electron neutrino mass is shifted at much higher
value then the neutrinos of the other two flavors. The mass spectrum is then formed by  a singlet eigenstate,
corresponding to pure electron neutrinos, and an eigenstate doublet below, involving the other two flavors.
During the neutrino propagation outwards, the electron density decrease and the mass difference between the
singlet and the doublet decreases, but to the extent that the mass difference stays large enough, the structure
of the spectrum remains unaltered. Since, at least in the so called "direct hierarchy" scheme, the lowest
neutrino mass in free space has a large electron flavor content, at a certain electron density the singlet
eigenstate must cross the doublet and strong mixing can occur, i.e. neutrinos can jump among the different local
(instantaneous) mass eigenstates, thus altering the neutrino populations among them. This is the essence of the
MSW mechanism. According to the degree of adiabaticity of the process, some flavor conversion can occur, which
determines the final flavor distribution. If the crossing is fast enough, the neutrinos in the initial electron
flavor singlet eigenstate will jump in the lowest eigenstate, which will finally merges in the lowest neutrino
mass in free space. The latter has a definite flavor content (mainly a mixture of electron and muon flavors). On
the contrary, if the process is completely adiabatic, the neutrinos will remain in the same (higher) local mass
eigenstate, and finally they will propagate in the highest mass eigenstate in free space (mainly a mixture of
$\tau$ and $\mu$ flavors). \par From this simple considerations it is clear that the crucial parameter of the
process is the adibaticity and therefore the density profile, which fixes the local density slope, has a direct
influence on the final neutrino signal. It is this connection between the density profile and the final mass and
flavor populations that we want to exploit systematically in this paper.

\par

In a simplified description where only two mass eigenstates are considered most of the calculations cannot be
done analytically, and a fortiori in a more realistic description with all three flavors included only numerical
simulations are possible. Following the standard treatment, the initial flavor distribution can be evolved by a
$3 \times 3$ matrix. In the case of three neutrino the flavor eigenstate $\nu_e$, $\nu_\mu$ and $\nu_\tau$ and
mass eigenstate $\nu_1$, $\nu_2$ and $\nu_3$  are related through the linear relation

\begin{equation}\label{eq_1}
\left( {\begin{array}{*{20}c}
   {\nu _e }  \\
   {\nu _\mu  }  \\
   {\nu _\tau  }  \\
\end{array}} \right) = \left( {\begin{array}{*{20}c}
   {U_{e1} } & {U_{e2} } & {U_{e3} }  \\
   {U_{\mu 1} } & {U_{\mu 2} } & {U_{\mu 3} }  \\
   {U_{\tau 1} } & {U_{\tau 2} } & {U_{\tau 3} }  \\
\end{array}} \right)\left( {\begin{array}{*{20}c}
   {\nu _1 }  \\
   {\nu _2 }  \\
   {\nu _3 }  \\
\end{array}} \right).
\end{equation}

\medskip

\medskip

In general, in the case of Dirac neutrinos the lepton mixing matrix
$U$ in eq.~(\ref{eq_1}) depends on three mixing angles
$\theta_{12}$, $\theta_{13}$ and $\theta_{23}$ and one CP-violating
phase~$\delta$, while in the case of Majorana neutrinos there are
two additional, so-called Majorana phases. It is convenient to use
the parametrization of the matrix $U$ which coincides with the
standard parametrization of the quark mixing matrix

\begin{equation}\label{eq_2}
U = \left( {\begin{array}{*{20}c}
   {c_{12} c_{13} } & {s_{12} c_{13} } & {s_{13} e^{ - i\delta } }  \\
   { - s_{12} c_{23}  - c_{12} s_{23} s_{13} e^{i\delta } } & {c_{12} c_{23}  - s_{12} s_{23} s_{13} e^{i\delta } } & {s_{23} c_{13} }  \\
   {s_{12} s_{23}  - c_{12} c_{23} s_{13} e^{i\delta } } & { - c_{12} s_{23}  - s_{12} c_{23} s_{13} e^{i\delta } } & {c_{23} c_{13} }  \\
\end{array}} \right)
\end{equation}

\medskip

\medskip

where $c_{ij} = \cos \theta_{ij}$, $s_{ij} = \sin \theta_{ij}$. The probabilities of oscillations between
various flavor states, unlike in the two-flavor case, are in general do not have a simple form. The evolution
equation describing neutrino oscillations in matter is

\begin{equation}\label{eq_3}
i\frac{d} {{dr}}\left( {\begin{array}{*{20}c}
   {\nu _e }  \\
   {\nu _\mu  }  \\
   {\nu _\tau  }  \\

 \end{array} } \right) = \left[ {\frac{1}
{{2E}}U\left( {\begin{array}{*{20}c}
   {m_1^2 } & 0 & 0  \\
   0 & {m_2^2 } & 0  \\
   0 & 0 & {m_3^2 }  \\

 \end{array} } \right)} \right.U^\dag   + \left. {\left( {\begin{array}{*{20}c}
   {V_{cc} } & \:\:0 & \:\:\:0  \\
   0 & \:\:0 & \:\:\:0  \\
   0 & \:\:0 & \:\:\:0  \\

 \end{array} } \right)} \right]\left( {\begin{array}{*{20}c}
   {\nu _e }  \\
   {\nu _\mu  }  \\
   {\nu _\tau  }  \\

 \end{array} } \right).
\end{equation}

\medskip

\medskip

Here $U$ is the $3 \times 3$ flavor vacuum mixing matrix defined in Eq. (\ref{eq_1}) and $V_{cc}(r)$ is the
matter-induced neutrino potential for unpolarized medium of zero total momentum defined in terms of electron
number density $N_e(r)$ and Fermi constant~$G_F$ as

\begin{equation}\label{eq_4}
V_{cc}(r)  = \sqrt 2 G_F N_e(r).
\end{equation}

It is rather difficult to study this equation analytically, even in
an adiabatic approximation. As we will see, the numerical treatment
is necessary also because in some cases the density profile does not
produce a real MSW resonance, but rather a ``modulation" of the
flavor populations, which can be hardly treated analytically. This
requires a quite stable numerical algorithm in solving iteratively
eq. (\ref{eq_3}). We used a method, based on the Hamilton-Caley
theorem, which guarantees the fulfillment of unitary to a very high
degree of accuracy and it is therefore suitable for calculations
which extend for large time an long distances. Typically the
calculations were performed from the center to the stars
up~to~$10^7$~kilometers.

Contrary to other methods used to solve the evolution equations of
the neutrino states by means of techniques like Runge-Kutta or
similar our numerical algorithm using the Hamilton-Caley's theorem
assures unitarity of the evolution operator to a very high degree of
accuracy. If the wave function of the flavor triplet of a neutrino
is written as a three components vector $\psi _\nu   \equiv \left(
{\nu _e ,{\rm{ }}\nu _\mu  ,{\rm{ }}\nu _\tau  } \right)^T$ the
flavor evolutions is completely determined by the Schr\"odinger
equation in the compact form

\begin{equation}\label{eq_5}i\frac{d}{{dr}}\psi _\nu(r)   = H\psi _\nu(r)\end{equation}

where the matrix $H$ is the Hamiltonian operator. The code obtain $ \psi _\nu  (l + \delta l)$ from $ \psi _\nu
(l)$ using an iterative procedure step by step:

\begin{equation}\label{eq_6}
\psi _\nu  (l + \delta l) \simeq e^{ - iH\delta l} \psi _\nu (l)
\end{equation}

preserving the unitary of $\psi _\nu$ automatically and becoming
exact if $H$ is indipendent of spatial coordinate. Therefore the
step sizes employed in numerical codes are not restricted by the
size of $H$ but are restricted by the rate of change of $H$. In this
algorithm the Hamilton-Caley's theorem is used to calculate step by
step the exponential operator in the eq. (\ref{eq_6}) to determinate
the evolution of $ \psi_\nu$ both in the simple two flavors
framework and in the standard three flavor framework.

\section{Neutrino signals and the protoneutron star structure.}

Numerical simulations of the collapse and evolution of supernovae up to the stage of protoneutron star is quite
difficult and affected by many uncertainties. For these reason simulations up to few tens of seconds after
collapse are quite rare. The reference simulations are those of the Livermore group \cite{wilson}, which are
often used in the literature for different types of analysis. These simulations include also the shock wave
front. Phenomenological models are also used for qualitative considerations \cite{fogli}.\par At the
protoneutron star stage of the evolution neutrinos are essentially thermally produced. If all three flavors had
the same temperature at the moment of decoupling from matter, i.e. the neutrino sphere is independent of flavor,
then all three flavors had the same fraction of $1/3$ and the same energy spectrum. Any possible conversion
mechanism would not affect such a population distribution and no characteristic neutrino signal could be
present. It is therefore essential to specify the initial temperature and chemical potential for each flavor in
the core of the star. For definiteness we will follow ref. \cite{fuller2}, but with the possibility of slightly
modify the parameters to check the sensitivity of the results on their specific values. Each flavor population
is described by a Fermi distribution

\par

\begin{equation}\label{eq_7}
f_\nu  (E) \equiv \frac{1}{{F_2 (\eta _\nu  )}}\frac{1}{{T_\nu ^3
}}\frac{{E^2 }}{{\exp (E/T_\nu   - \eta _\nu  ) + 1}},
\end{equation}

\noindent

\medskip

\medskip

 where $\eta_\nu$  is the degeneracy parameter, $T_\nu$ is the neutrino temperature and

\begin{equation}\label{eq_8}
F_k (\eta ) \equiv \int_0^\infty  {\frac{{x^k dx}}{{\exp (x - \eta )
+ 1}}}
\end{equation}

\noindent
is the Fermi integral with rank $k=2$.

The values of the degeneracy parameters are taken $\eta _{\nu _e } = \eta _{\nu _\mu  }  = \eta _{\nu _\tau  } =
3$ and neutrino temperatures are taken $ T_{\nu _e } \simeq 2.76~{\rm{ MeV}}$ and $T_{\nu _\mu  }  = T_{\nu
_\tau  } \simeq 6.26~{\rm{ MeV}}$. The normalization factors $F_2 (\eta _\nu  )$ specify also the total content
of each flavor. Once the initial conditions are fixed, one has to define the parameters of the flavor mixing $3
\times 3$ matrix. Presently the only parameter that remain uncertain is the so-called $\theta_{13}$ mixing
angle, while for all the others, standard established values can be used \cite{bal}. We choose the value $\sin^2
\theta_{13} \, =\, 10^{-4}$. \par
\begin{figure}[h]
\begin{center}
\includegraphics[width=1.0\textwidth]{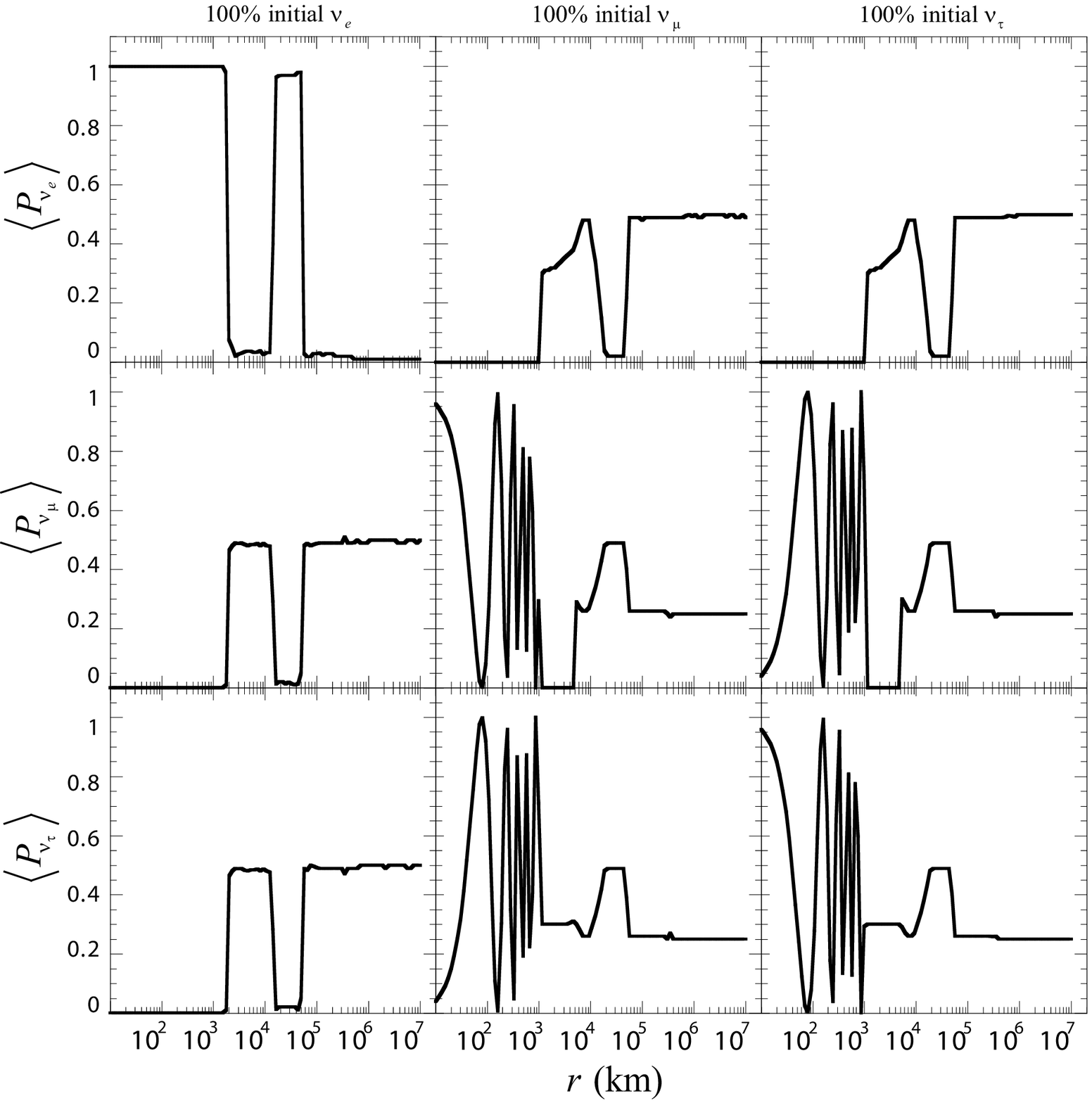}
\caption{Survival probabilities and transition probabilities for different flavors along neutrino path inside
the protoneutron star matter.
 The energy of neutrinos is 3 MeV and the time after bounce is t = 4 s. The density profile is
 characterized by the parameter $\alpha$ = 2.1, see the text for detail} \label{figure1}
\end{center}
\end{figure}

 Since the density profile changes with time also the final neutrino signal is
expected to change with time. For illustration, at the fixed time of $ t = 4~{\text{s}} $ and for the energy $ E
= 3~{\text{MeV}} $ the flavor survival probabilities are reported in~Fig.~\ref{figure1} as a function of the
distance from the core of the star up to radius of $10^7$ kilometers, where the influence of matter is surely
not relevant. For this reason the probabilities or transition probabilities are usually rapidly oscillating in
the final part of the propagation. We have then followed an averaging procedure over many oscillation period to
establish the final average value of each flavor content. This is justified by the wide region where neutrino
are actually emitted, which introduces a random phase. In any case, one can easily recognize the place where the
MSW resonances occur.

This actually corresponds to the $higher$ resonance, i.e. the place where the initial highest mass eigenvalue
crosses the first one below.

\par

\begin{figure}[h]
\begin{center}
\includegraphics[width=1.0\textwidth]{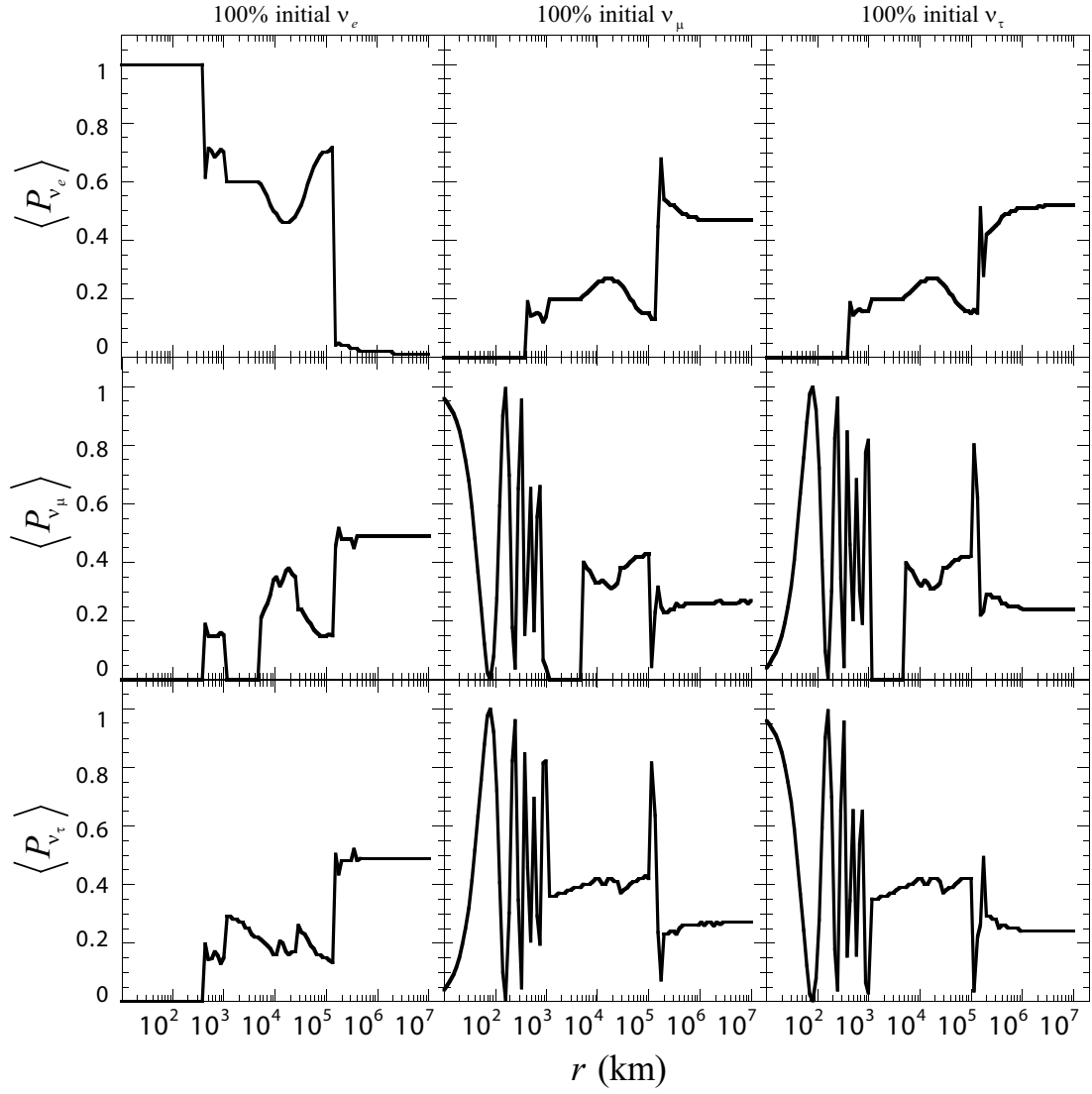}
\caption{The same as in Fig. 1 but at the different time t = 12 s.} \label{figure2}
\end{center}
\end{figure}

In each row is reported the survival probability of each flavor
(along the diagonal) and the contribution to each flavor coming from
the other two flavors (off diagonal panels). In other words each
panels contains the evolution of the flavor conversion probabilities
along the neutrino path. At each radial distance $r$ the sum of the
values reported in a given row gives the total survival probability
of each flavor. It has to be noticed the relevance of the
``re-population" of e.g. the electron flavor by the other two
flavors, without which the electron flavor content would drop to
zero. Another example with different values of energy and fixed time
is in Fig. \ref{figure2}. Of course the density profile changes
during the neutrino propagation but the main conversion processes
occur in a time interval short enough that the density profile can
be considered fixed. In this sense the nominal value of time $t(s)$
must be taken just as a nominal \mbox{value labeling the particular
case considered}.

From figures 1 and 2 it is apparent the interplay of the MSW conversion process and the non-resonant modulation
due to the density profile. This is mainly due to the fact that the MSW resonance width is quite wide, and some
degree of flavor conversion can occur even if the resonance density is not strictly crossed. The final signal
depends in a complex way on the density profile. This also indicates the need of numerical calculations. It has
to be stressed that the density at which the MSW resonance is expected to occur is quite low with respect to the
typical nuclear density and in this density region matter is dominated by electrons, positrons and photons.
However, the overall density profile depends also indirectly on the nuclear Equation of State (EOS), since the
EOS is one of the main physical ingredient which determines the evolution and structure of the protoneutron
star, and thus the density profile even in its outer part.

\par

\begin{figure}[h]
\begin{center}
\includegraphics[width=1.0\textwidth]{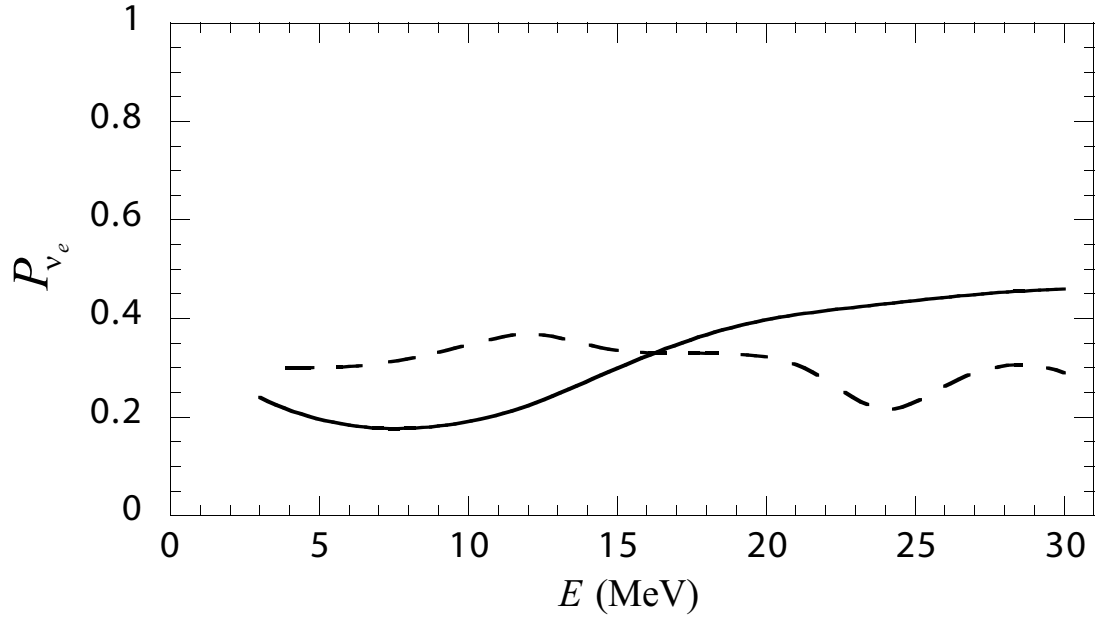}
\caption{Fraction of electron neutrinos as a function of energy at t = 4 s (full line) and t = 27 s (dashed
line) after bounce. The parameter of the profiles has been fixed at $\alpha$ = 2.4 .} \label{figure3}
\end{center}
\end{figure}

We have repeated these calculations at different times and different neutrino energies. The final neutrino
energy spectrum of each flavor can be then extracted at different protoneutron star evolution time. This can be
obtained by multiplying the initial Fermi distributions by the survival probability.  In Fig. \ref{figure3} is
reported the fraction of electron neutrino for two different times as a function of neutrino energy. It is clear
that in these cases the electron neutrino fraction is only slightly modulated as a function of time. Also the
initial spectrum is affected by the propagation through the protoneutron star matter, however the final spectrum
is expected to change only slightly at different times.

\begin{figure}[h]
\begin{center}
\includegraphics[width=1.0\textwidth]{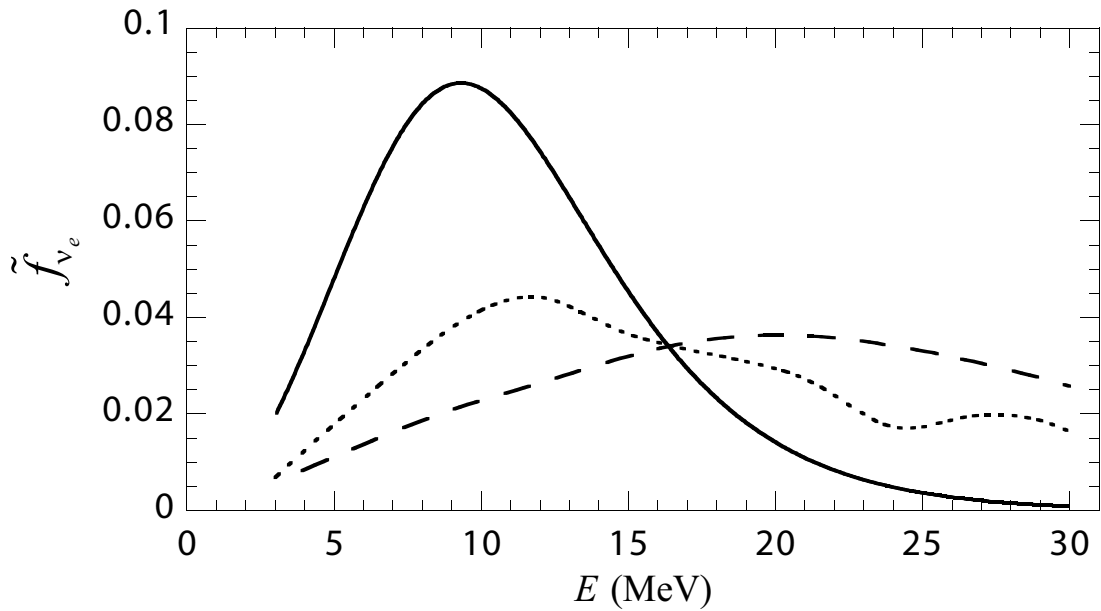}
\caption{electron neutrino spectrum corresponding to the case of Fig. 3. The full line is the initial
Fermi-Dirac spectrum, the dashed line is the spectrum at t = 4 s and the dotted line at t = 27 s.}
\label{figure4}
\end{center}
\end{figure}

Indeed the spectrum is reported in Fig. (\ref{figure4}) at the two different times in comparison with the
initial one. It has to be noticed the strong distortion of the initial spectrum, due to the conversion process,
but this distortion depends little on time. Of course the overall normalization depends on the total neutrino
flux, but the considered properties, i.e. the electron neutrino fraction and the corresponding spectrum, do not
depend on the value of the flux.

To see the sensitivity of these results on the density profile we have repeated the calculations by changing
artificially the density profile of the protoneutron stars. To this purpose we have followed the profile
parametrization introduced in ref. \cite{fogli}, according to which the density profile can be effectively
approximated for post-bounce times $ t \lesssim 1{\text{ s}}$ as

\begin{equation}\label{eq_9}
\frac{{\rho _0 (r)}}{{{\rm{g/cm}}^{\rm{3}} }} \approx 10^{14} \left(
{\frac{r}{{{\rm{Km}}}}} \right)^{ - \alpha }
\end{equation}

\noindent

\medskip

\medskip

 and for post-bounce time $ t \gtrsim 1{\text{ s}} $ as

\begin{equation}\label{eq_10}
\rho (r) = \rho _0 (r) \cdot \left\{ \begin{gathered}
  \xi  \cdot f(r)\quad,\quad r \leqslant r_S \quad , \hfill \\
  1\quad \quad \quad \; \, \, ,\quad r \geqslant r_S \quad , \hfill \\
\end{gathered}  \right.
\end{equation}

%\noindent
 where $ \xi  \simeq 10 $ and the function $f(r)$ parametrizes the rarefaction zone above the shock
front. The parametrization for $f(r)$ is taken

\begin{equation}\label{eq_11}
\ln f(r) = [0.28 - 0.69\ln (r_S /{\text{Km}})] \cdot [\arcsin (1 -
r/r_S )]^{1.1}.
\end{equation}

\par\noindent

In eq. (\ref{eq_10}) and eq. (\ref{eq_11}) it is assumed a slightly
accelerating shock-front position~$x_S$,

\begin{equation}\label{eq_12}
r_S (t) = r_S^0  + v_S t + \frac{1} {2}a_S t^2
\end{equation}

\noindent
with parameters fixed by

\begin{equation}\label{eq_13}
\begin{gathered}
  x_S  \simeq  - 4.6 \times 10^3 {\text{km }}{\text{,}} \hfill \\
  v_S  \simeq 11.3 \times 10^3 {\text{km/s }}{\text{,}} \hfill \\
  a_S  \simeq 0.2 \times 10^3 {\text{km/s}}^{\text{2}} , \hfill \\
\end{gathered}
\end{equation}

\par
\begin{figure}[h]
\begin{center}
\includegraphics[width=1.0\textwidth]{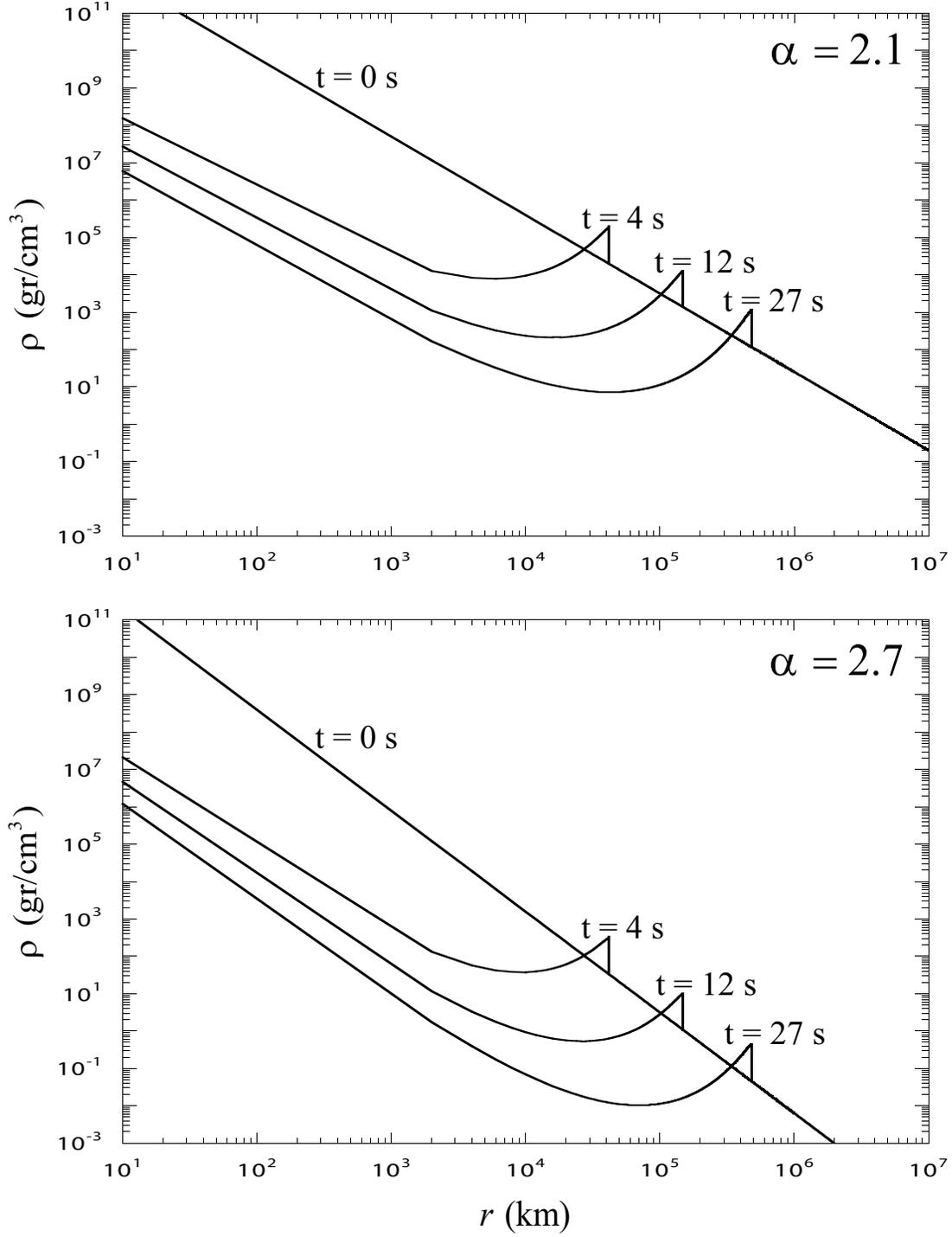}
\caption{Density profiles at different times after bounce corresponding to $\alpha$ = 2.1 (left) and to $\alpha$
= 2.7 (right).} \label{figure5}
\end{center}
\end{figure}

\noindent
The value of the parameter $\alpha = 2.4$ corresponds to the parametrization which reproduce closely the
simulations of ref. \cite{wilson}, and this is the value used in the calculations reported above. We then varied
the value of $\alpha$ to obtain a steeper or a more shallow density profile, which should possibly reflect
different model simulations of the proton-neutron star dynamics. This is illustrated in Fig.~\ref{figure5}~,
where the density profiles for different values of $\alpha$ are compared at fixed times.

\begin{figure}[h]
\begin{center}
\includegraphics[width=1.0\textwidth]{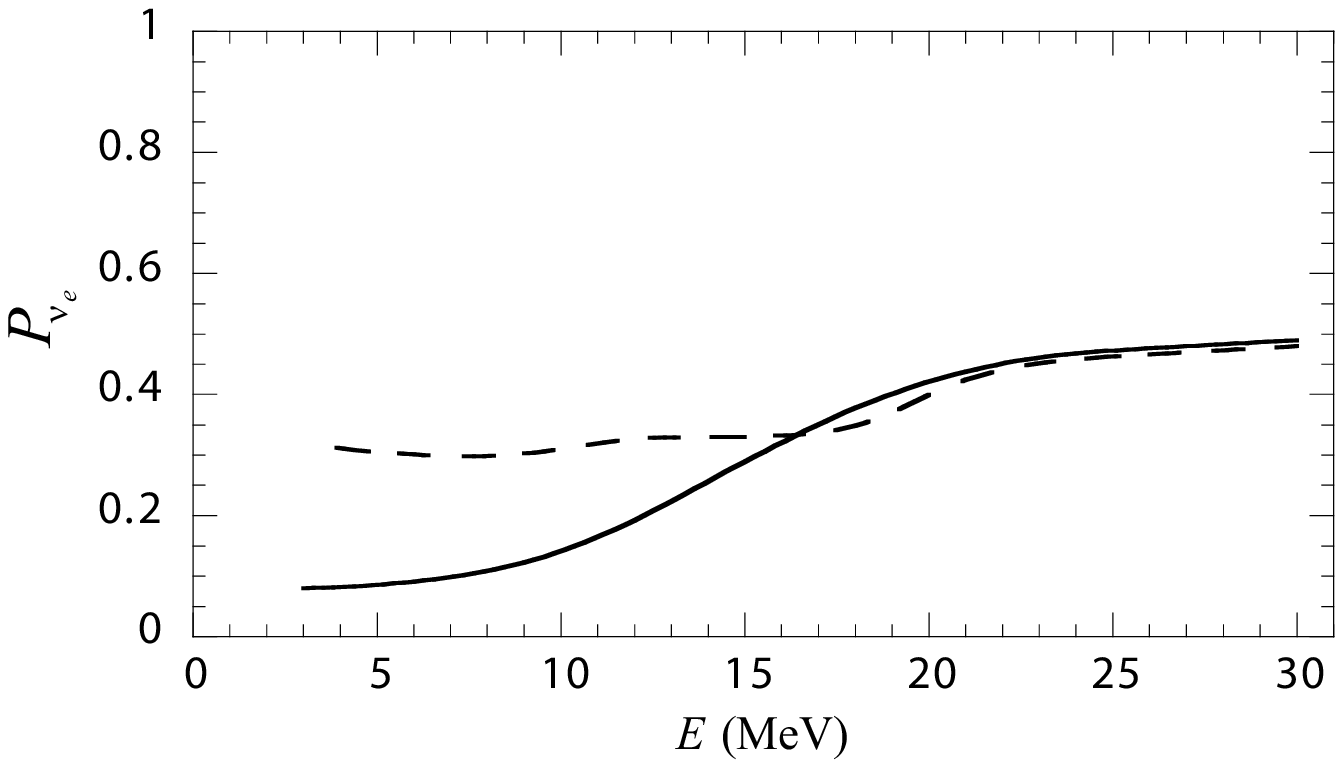}
\caption{The same as in Fig. 3 but for $\alpha$ = 2.1 .} \label{figure6}
\end{center}
\end{figure}

\begin{figure}[h]
\begin{center}
\includegraphics[width=1.0\textwidth]{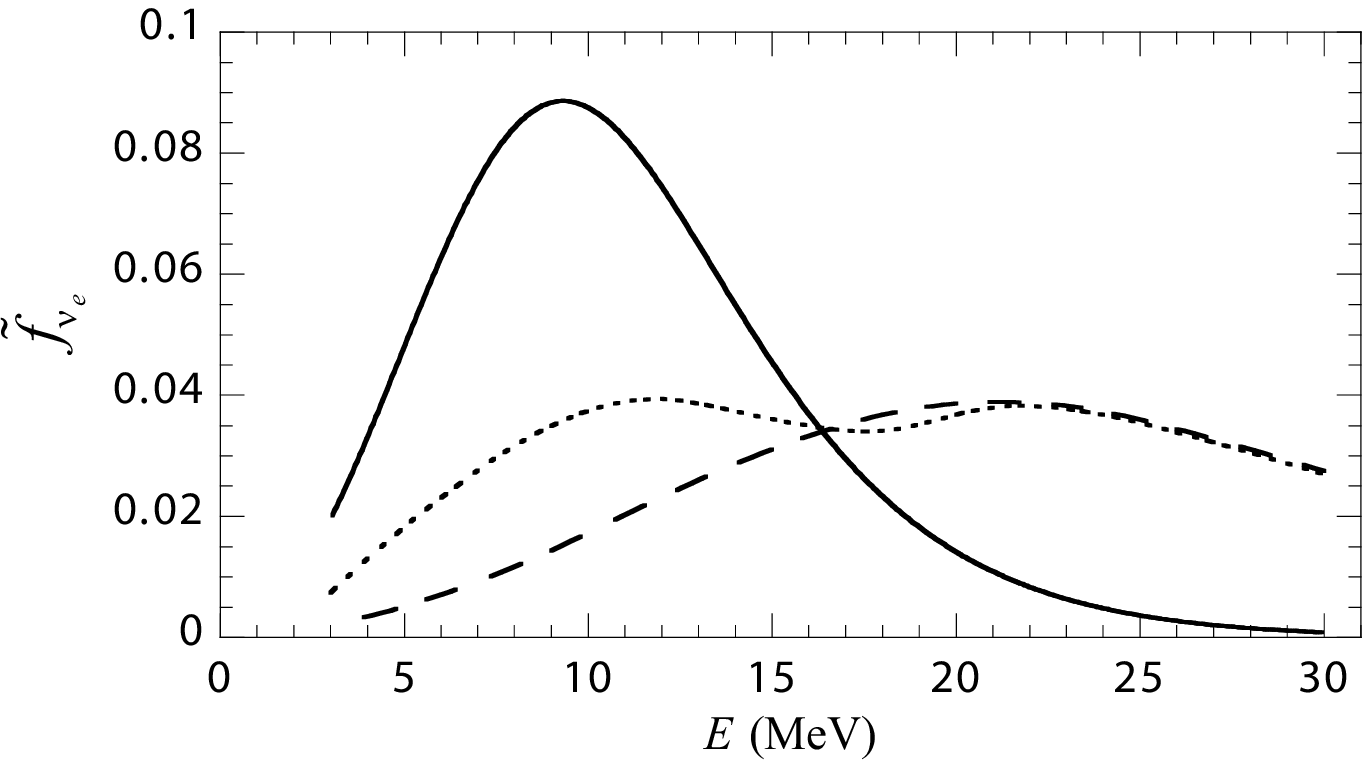}
\caption{The same as in Fig. 4 but foe $\alpha$ = 2.1 .} \label{figure7}
\end{center}
\end{figure}

 The modification of the profile has a direct effect on the neutrino fraction as shown in Fig. \ref{figure6}
for the case of $\alpha = 2.1$. At the lower energy, below about 15
MeV, the electron neutrino fraction presents a sharp drop at an
early time evolution of the protoneutron star. This means that the
electron neutrino energy spectrum should be drastically suppressed
in the low energy part. At a later stage the energy spectrum should
change again. The final spectra at the two different times, in
comparison with the initial Fermi-Dirac spectrum, for electron
neutrinos are reported in Fig. \ref{figure7}. Also in this case the
final spectrum is strongly distorted by the conversion process which
occurs during the neutrino propagation in the protoneutron star
matter, but what is relevant is that the distortion depends now more
clearly on time. The increase of the high energy tail is due to the
conversion of $\mu$ and $\tau$ neutrinos in electron neutrinos.
Modification of the initial Fermi-Dirac distribution can also occur
due to energy dependence of the neutrino transparency \cite{janka},
but the type of dynamical evolution of the spectrum reported in Fig.
7 should be a distinct feature of a shallow density profile, also
independent of the total neutrino flux, which also depends on time.
Comparison between Figs. \ref{figure3},\ref{figure4} and Figs.
\ref{figure6},\ref{figure7} shows the sensitivity on the density
profile of the protoneutron stars during its evolution.

The reason for such a behavior is directly linked to the higher
degree of adiabaticity of the MSW mechanism, in the wider sense
discussed above. In fact, increasing the value of $\alpha$ to 2.7
both electron neutrino fraction and spectrum become almost
independent of time, see Figs. 8,9.

%The neutrino survival probability for the case of $\alpha =
%2.7$ is shown in Fig. (\ref{figure6}).
We have also tested the dependence of the results on the assumed values of the neutrino temperatures. Decreasing
the temperature differences all the considered effects decrease, which indicate the possibility of measuring the
original neutrino temperatures.

\section{Conclusions.}

In this paper we have presented evidence of possible neutrino
signals from protoneutron stars based on the flavor conversion
processes which can occur inside the star and which are sensitive to
the density profile of these astrophysical compact objects.
Generally speaking a more shallow density profile would produce a
percentage of electron neutrino strongly suppressed at low energy,
approximately below 15 MeV, few second after collapse. This
suppression should disappear after 10-20 seconds. Correspondingly,
the electron neutrino spectrum should show a characteristic time
dependence. \par Of course, all these signals require an appreciable
number of events in the detectors, and only with the next generation
of neutrinos detectors one can hope to have indications of the
signals studied in the present paper.

%\begin{figure}
%\begin{center}
%\includegraphics[width=1.0\textwidth]{figure6.eps}
%\caption{Didascalia da scivere.} \label{figure6}
%\end{center}
%\end{figure}

%\clearpage

\par

\begin{figure}[h]
\begin{center}
\includegraphics[width=1.0\textwidth]{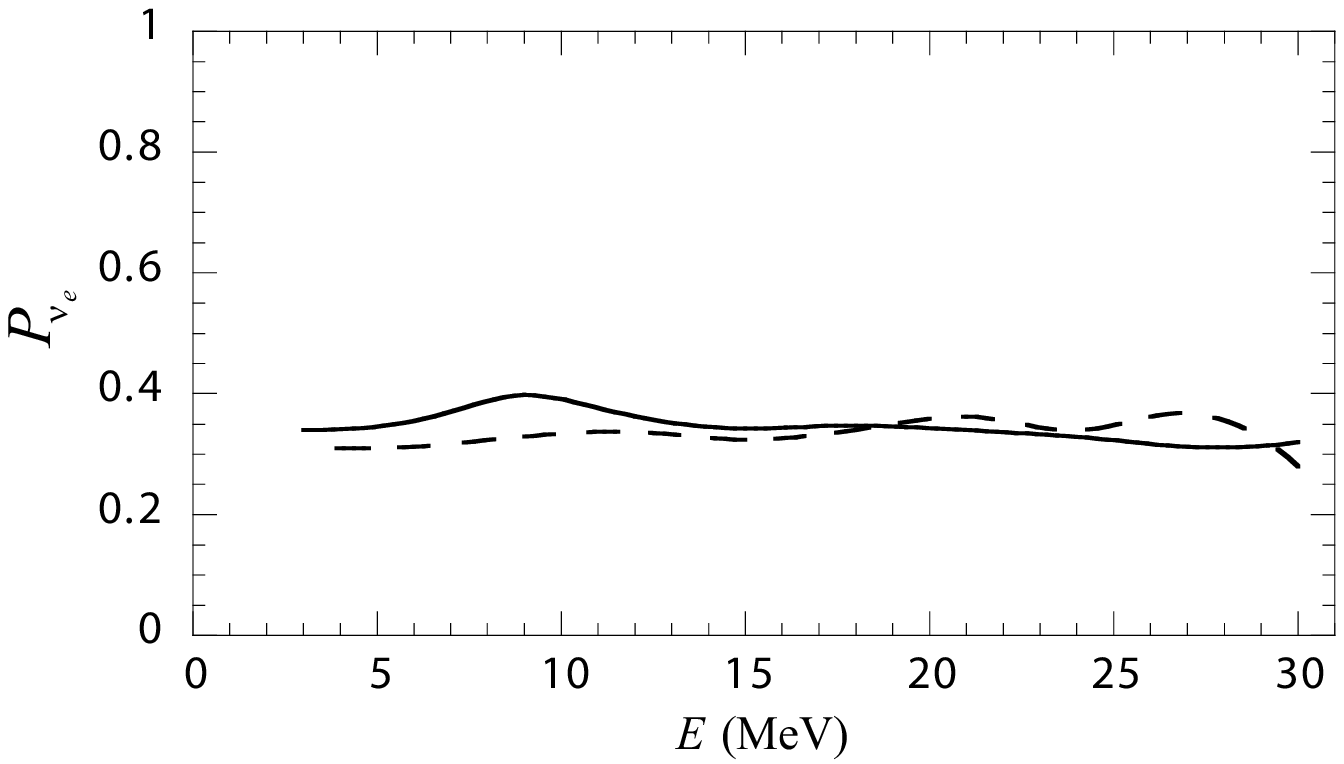}
\caption{The same as in Fig. 3 but for $\alpha$ = 2.7 .} \label{figure8}
\end{center}
\end{figure}

\par

\begin{figure}[h]
\begin{center}
\includegraphics[width=1.0\textwidth]{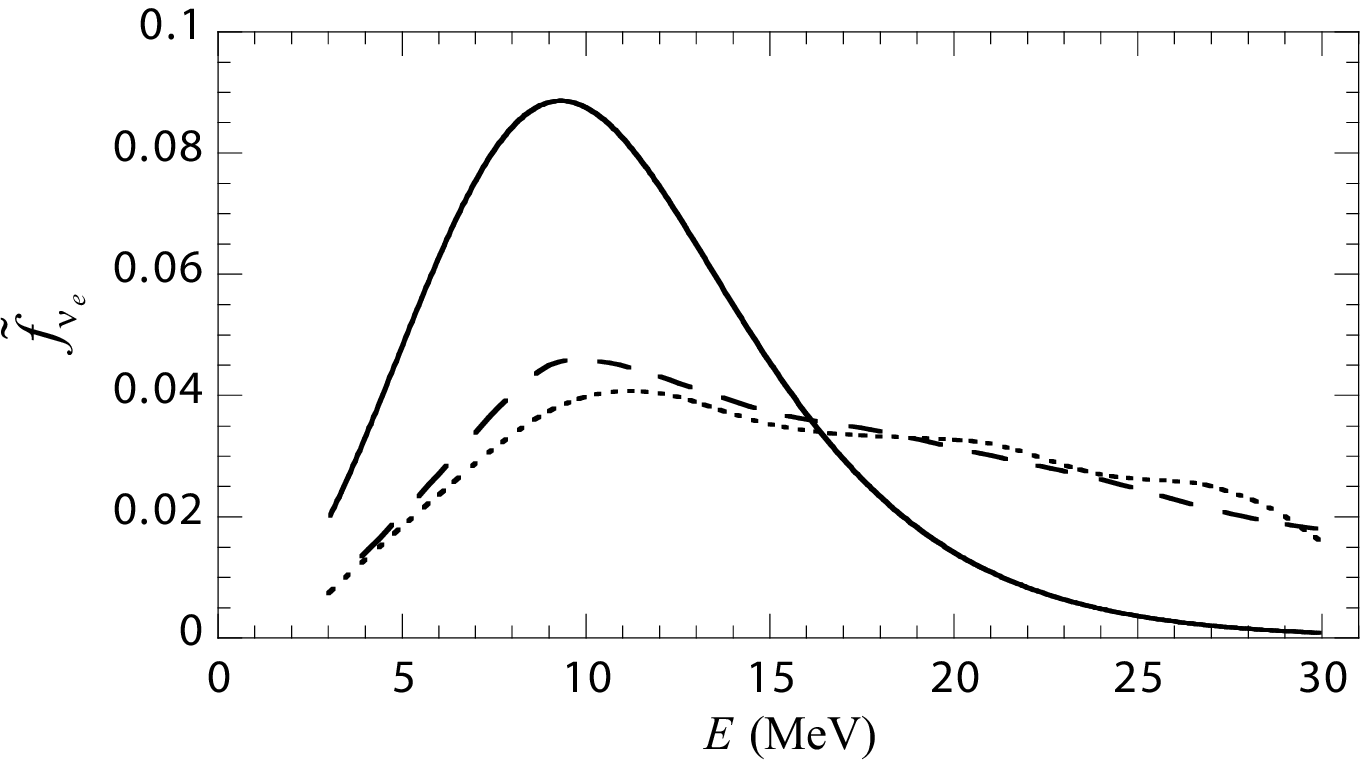}
\caption{The same as in Fig. 4 but for $\alpha$ = 2.7 .} \label{figure9}
\end{center}
\end{figure}

\end{document}